\documentclass[preprint,pra,superscriptaddress]{revtex4-1}

\usepackage[dvips]{graphicx} % for figures
\usepackage{amsfonts}
\usepackage{amssymb}
\usepackage{amscd}
\usepackage{amsmath}    % need for subequations
\usepackage{enumerate}
\usepackage{epsfig}
\usepackage{subfigure}
\usepackage{xcolor}

\begin{document}

% title
\title{Statistical fluctuation analysis for measurement-device-independent quantum key distribution}

\author{Xiongfeng Ma}
\email{xma@tsinghua.edu.cn}
\affiliation{Center for Quantum Information, Institute for Interdisciplinary Information Sciences, Tsinghua University, Beijing, China}

\author{Chi-Hang Fred Fung}
\affiliation{Department of Physics and Center of Theoretical and Computational Physics, University of Hong Kong, Pokfulam Road, Hong Kong}%

\author{Mohsen Razavi}
\affiliation{School of Electronic and Electrical Engineering, University of Leeds, Leeds, United Kingdom}

%%%%%%%%%%%%%%%%%%%%%%%%%%%%%%%%%%%%%%%%%%%%%%%%%%%%%%%%%%%%%%%%%%%%%%%
% Abstract
%%%%%%%%%%%%%%%%%%%%%%%%%%%%%%%%%%%%%%%%%%%%%%%%%%%%%%%%%%%%%%%%%%%%%%%
\begin{abstract}
Measurement-device-independent quantum key distribution with a finite number of decoy states is analyzed under finite-data-size assumption. By accounting for statistical fluctuations in parameter estimation, we investigate vacuum+weak- and vacuum+two-weak-decoy-state protocols. In each case, we find proper operation regimes, where the performance of our system is comparable to the asymptotic case for which the key size and the number of decoy states approach infinity. Our results show that practical implementations of this scheme can be both secure and efficient.
%enable the secure implementation of this scheme in practice.
\end{abstract}

\maketitle

\section{Introduction}
Quantum key distribution (QKD)~\cite{BB_84,Ekert_91} is one of the most successful applications of quantum information processing, which allows two distant parties, Alice and Bob, to grow secret keys with information-theoretic security \cite{Mayers_01,LoChauQKD_99,ShorPreskill_00,Biham:SecurityPf:2000,ILM_07,Renner:SecurityPf:2005}. Conventional security proofs of QKD assume certain physical models for the employed devices --- source and detection units. For instance, the squashing model is widely assumed for the measurement \cite{BML_Squash_08,TT_Thres_08,Fung:2011:Squash} in a standard security analysis \cite{GLLP_04}. Practical implementations, however, could fall short of meeting all requirements set by the models, hence security could be compromised in reality. In fact, side channels have been identified and exploited to break QKD security. These side-channel attacks include the fake-state attack~\cite{MAS_Eff_06,Makarov:Fake:08}, the time-shift attack~\cite{Qi:TimeShift:2007,Zhao:TimeshiftExp:2008}, the phase-remapping attack~\cite{Fung:Remap:07,Xu:PhaseRemap:2010}, and the detector-blinding attack~\cite{Lydersen:Hacking:2010,Gerhardt:Blind:2011}.

Several approaches have been proposed to counter the side-channel attacks. One way is to sufficiently characterize the behavior of the devices and analyze the security by taking into account all device parameters \cite{Fung:Mismatch:2009,Lydersen:Mismatch:2010,Maroy:imperfection:2010}. This, however, can be difficult to implement in practice. A second approach that can defeat all side-channel attacks is device-independent QKD~\cite{MayersYao_98,Acin:DeviceIn:07,Pironio:DeviceIn:09}, in which the security can be proven without knowing the specifications of the devices used. Security, in this case, is derived from nonlocal correlations by violating Bell's inequality \cite{Bell_Ineq_64,CHSH.Bell.69}. In order to avoid the detection efficiency loophole~\cite{Pearle_Bell_70},  however, a large fraction of the transmitted signals must be detected by the receiver, resulting in impractical requirements for the transmission efficiency (e.g., $82.8\%$~\cite{GM_Bell_87} for the Clauser-Horne-Shimony-Holt (CHSH) inequality \cite{CHSH.Bell.69}).

Instead of full device independence, a detection-device independent QKD scheme is proposed \cite{MXF:DDIQKD:2012,Pawlowski:Semi:2011}, in which the detection system is assumed to be untrusted. Since most of practical hacking strategies focus on the detection site, and the source site is relatively simple for characterization, such a scheme can close most loopholes in a QKD system. Unfortunately, these schemes still need stringent requirements on the transmission efficiency of more than 50\% \cite{MXF:DDIQKD:2012}.

Recently, Lo, Curty, and Qi~\cite{Lo:MIQKD:2012} proposed efficient schemes that are measurement-device independent (MDI). Alice and Bob send some signals to a willing participant who can even be an eavesdropper, Eve. Eve performs a Bell-state measurement (BSM) and announces the result to Alice and Bob who will use this information to distill a secret key. The security is based on the idea of entanglement swapping using a BSM and the reverse EPR QKD scheme \cite{Biham:ReverseEPR:1996,Inamori:Security:2002,Braunstein:MIQKD:2012}. The scheme is secure even if Eve intentionally makes the wrong measurement and/or announces the wrong information. Various implementation approaches to MDI-QKD have also been proposed  \cite{Tamaki:MIQKD:2012,MXF:MIQKD:2012}, and significant efforts have been devoted to its experimental demonstration \cite{Lo:MIQKD:2012,Rubenok:MIQKDexp:2012,daSilva:MIQKD:2012}. Recently, the first MDI-QKD experiment with decoy states is completed by Liu \emph{et al.}~\cite{Liu:MIQKDexp:2012}.

%~\cite{Lo:MIQKD:2012,Braunstein:MIQKD:2012}

MDI-QKD is not completely device independent and the source devices have to be trusted and sufficiently characterized. When we use a coherent source to implement a single-photon-based MDI-QKD scheme, such as that in Ref.~\cite{Lo:MIQKD:2012}, we need to estimate the single-photon contributions of the detection at the receiver, which can be done efficiently using decoy states \cite{Hwang:Decoy:2003,Lo:Decoy:2005,MXF:Practical:2005,Wang:Decoy:2005,*Wang:Decoy2:2005}. In \cite{Lo:MIQKD:2012}, a security analysis is provided for the decoy-state MDI-QKD assuming infinitely long keys with infinitely many decoy states. In this paper, we proceed further and analyze the performance of decoy-state MDI-QKD when only a finite number of decoy states are used. Moreover, we consider statistical fluctuations caused by a finite-size key. Such an analysis is crucial to ensure the security of MDI-QKD in practical setups.

%Our analysis of decoy states with statistical fluctuation is quite general and is applicable to both polarization- and phase-encoding MDI-QKD schemes. We demonstrate our idea on a BB84 based MDI-QKD scheme that uses phase encoding.

We note that the effect of finite size on MDI-QKD has also been recently studied in an independent work by Song {\it et al.}~\cite{Song:2012:finiteMDIQKD}.
However, they only analyzed the vacuum+weak-decoy-state protocol whereas we also analyze the vacuum+two-weak-decoy-state protocol here taking advantage of our general method which can easily be adapted to other decoy-state protocols.

The rest of this paper is organized as follows. In Sec.~\ref{Sec:MIFluc:MIQKDrev}, we briefly review the MDI-QKD scheme with decoy states. In Sec.~\ref{Sec:MIFluc:Model}, we investigate the QKD model for the security proof and simulation. In Sec.~\ref{Sec:MIFluc:PostProcessing}, we perform a statistical fluctuation analysis on MDI-QKD systems, followed by numerical results in Sec.~\ref{Sec:MIfluc:Sim}. We conclude the paper in Sec.~\ref{Sec:MIFluc:Conclusion} with remarks.

\section{Decoy-state MDI-QKD} \label{Sec:MIFluc:MIQKDrev}
\begin{figure}[hbt]
\centering
\resizebox{12cm}{!}{\includegraphics{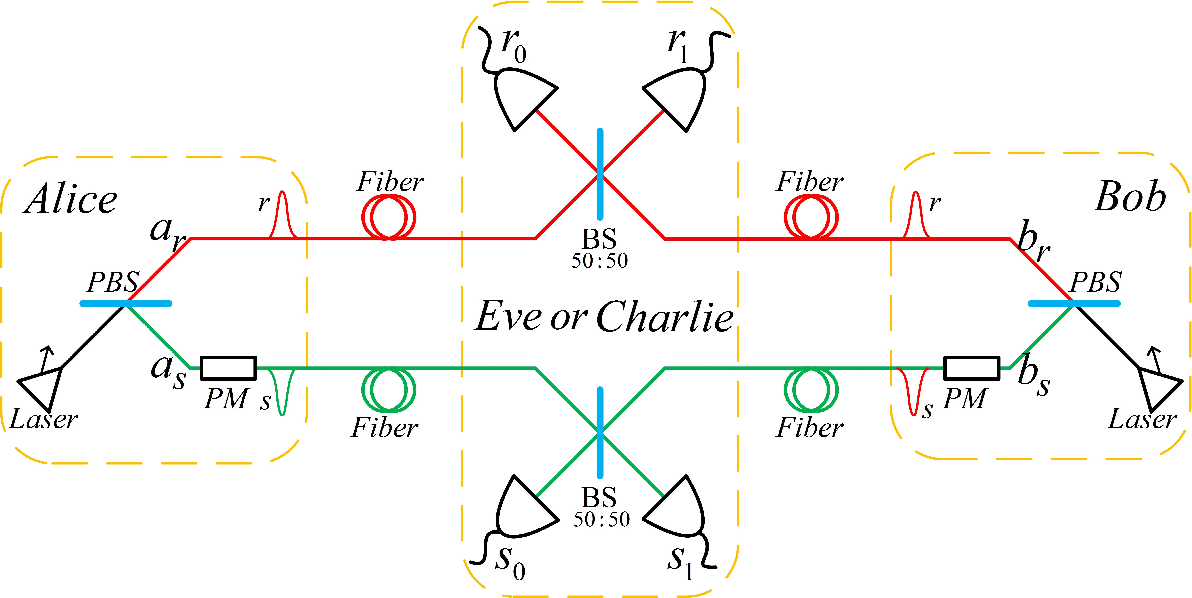}}
\caption{(Color Online) A schematic diagram for the MDI-QKD protocol, where PBS stands for polarizing beam splitter and PM stands for phase modulator. In order to encode their bits in the $z$ basis, Alice and Bob generate phase-randomized coherent states with either $H$ or $V$ polarizations at their sources. To encode a bit in the $x$ or $y$ basis, they generate $+45$-polarized signals at their encoders. A PM will introduce a relative phase shift between their reference and signal beams. The phase shifts are chosen from the set $\{ 0 , \pi \}$ for the $x$ basis and  $\{ \pi/2 , 3 \pi /2 \}$ for the $y$ basis. A partial BSM, possibly performed by an untrusted party, Eve or Charlie, on the two reference and the two signal modes would establish correlations between the raw key bits of Alice and Bob. If they both use the $z$ basis, a click on exactly one of the $r$ detectors {\em and} exactly one of the $s$ detectors would imply anti-correlated bits shared between Alice and Bob. For $x$ and $y$ bases, if they both use the same basis, a joint click on detectors $r_0$ and $s_0$ implies identical bits for Alice and Bob; so does a joint click on $r_1$ and $s_1$. A joint click on $r_0$ and $s_1$, or, $r_1$ and $s_0$ would imply anticorrelated bits \cite{MXF:MIQKD:2012}.}
\label{Fig:Path:MIdiag}
\end{figure}

The most general encoding scheme for BB84-based QKD relies on using two optical orthogonal modes. Here, we encode a qubit in the $z$ basis by using two spatially separated modes, $r$ and $s$, as shown in Fig.~\ref{Fig:Path:MIdiag}. That is, for the $z$ basis, the information is encoded in whether the photon is in mode $r$ or $s$. The qubit can also be encoded into the relative phases between modes $r$ and $s$. Denote $x$ basis to be the case when two relative phases $\{0,\pi\}$ are used and $y$ basis for $\{\pi/2,3\pi/2\}$. This encoding is sufficiently general to be tailored down to all proposed MDI-QKD schemes. For example, in the original MDI-QKD \cite{Lo:MIQKD:2012}, $r$ and $s$ correspond to $H$ and $V$ polarizations. For BB84 encoding, the $z$ and $x$ basis is used \cite{Lo:MIQKD:2012}. We remark that this setup can be used to implement the six-state QKD protocol as well \cite{Bruss:6state:1998}. For practical purposes, one may consider using temporal, rather than the spatial, modes as proposed in \cite{MXF:MIQKD:2012, Rubenok:MIQKDexp:2012}. Here, however, we are mostly concerned with statistical fluctuation effects due to the finite size of the key, and our results are independent of the employed setup. The key assumption in all MDI-QKD schemes is that the photons on which the intermediary BSM is performed are indistinguishable. We assume this condition is held throughout our analysis.

%This scheme is proposed in \cite{MXF:MIQKD:2012}. We note that the encoding is generic. The optical modes, $r$ and $s$, could be the polarizations, $H$ and $V$, which is used in the original MDIQKD scheme \cite{Lo:MIQKD:2012}.

In this paper, we assume that Alice and Bob use coherent states as their sources and use the $z$ and $x$ basis above for encoding. The MDI-QKD scheme runs as follows.

\begin{enumerate}
\item
Alice randomly chooses a basis from $\{x,z\}$ and a bit from $\{0,1\}$, and sends a coherent-state pulse with intensity randomly chosen from a predetermined set. As shown in Fig.~\ref{Fig:Path:MIdiag}, if she picks the $z$ basis, she prepares her coherent states with either $H$ or $V$ polarizations depending on the bit value. Otherwise, if she picks the $x$ basis, she prepares $+45$-polarized signals, splits the pulse into two modes, $r$ and $s$, through a polarizing beam splitter (PBS), and encodes the bit values into relative phases, $\{0,\pi\}$, between the two modes. Bob applies the same encoding procedure.

\item
Alice and Bob send the pulses to the relay, which can be fully controlled by Eve. Eve performs a partial BSM on the received pulses, as shown in Fig.~\ref{Fig:Path:MIdiag}. Eve announces her detection results. She is allowed to be dishonest.

\item
Alice and Bob compare the bases used for all transmissions which include the no-detection events, successful BSM events, and unsuccessful BSM events.

\item
Based on Eve's announcement for each pulse, Alice and Bob keep the bit if it corresponds to a successful BSM event and a compatible basis has been used.  One of them also flips the bit value in the case of an anticorrelated BSM result (see Fig.~\ref{Fig:Path:MIdiag}).
They discard all other bits corresponding to the no-detection events, unsuccessful BSM events, and those with incompatible bases.

%\item
%Based on Eve's announcement for each pulse, Alice and Bob keep the bit if it is a successful BSM and flip the bit value if needed.
%
%\item
%Alice and Bob compare the bases used for all bits regardless of whether successful BSMs have occurred, and keep the compatible bits.

\item
For each combination of Alice's intensity $\mu$, Bob's intensity $\nu$, and basis $w=x,z$, they test the error rate $E_{\mu\nu}^{w}$ of the retained bits, and compute the gain $Q_{\mu\nu}^{w}$ by counting the number of successful BSM events among all transmissions (including the no-detection events, successful BSM events, and unsuccessful BSM events) when Alice and Bob used compatible bases.
Thus, it is necessary for Alice and Bob to compare their bases even for bits that have not resulted in a successful BSM and are to be discarded.

\item
Alice and Bob estimate the yield $Y_{11}^{z}$ and the phase error rate $e_{11}^{x}$ for the fraction of signals in which Alice has a single photon and Bob has a single photon, based on the analysis in Sec.~\ref{Sec:MIFluc:PostProcessing}. With this parameter estimation, Alice and Bob perform error correction and privacy amplification to distill a final secret key.

\end{enumerate}

The analysis in the last step is the main focus of this work.

\section{Model} \label{Sec:MIFluc:Model}
The notations and definitions used in the model are listed below.
\begin{itemize}
\item
Alice and Bob each use coherent states to implement decoy-state MDI-QKD. In addition to the signal state, different intensities will be used for a number of decoy states. In this section, we denote the mean number of photons in a certain pulse sent by Alice and Bob, respectively, by $\mu$ and $\nu$. In subsequent sections, we introduce a more detailed notation as needed for decoy states.

\item
We use the term ``$i$-photon channel'' when a Fock state with $i$ photons is used as information carrier. We denote the joint channel when Alice uses an $i$-photon channel \emph{and} Bob uses a $j$-photon channel by $i\uplus j$ channel, where $i,j=0,1,2,\dots$. When there is no ambiguity, we use $\mu\uplus\nu$ channel to represent the case when Alice and Bob send out coherent states with intensities $\mu$ and $\nu$, respectively.

\item
The overall gain $Q_{\mu\nu}^{w}$ is defined as the probability of obtaining a successful partial BSM \emph{when} Alice and Bob use the $\mu\uplus\nu$ channel and the $w$ basis, where $w=x,z$. The quantum bit error rate (QBER) $E_{\mu\nu}^{w}$ is the corresponding error probability.

\item
The yield $Y_{ij}^{w}$ is the probability to obtain a successful BSM when Alice and Bob use the $i\uplus j$ channel and the $w$ basis, where $w=x,z$, and $e_{ij}^{w}$ is the corresponding error probability. The gain $Q_{ij}^{w}$ is defined as the probability that Alice and Bob use the $i\uplus j$ channel \emph{and} obtain a successful partial BSM.

\item
Denote the transmittance of the channel between Alice (Bob) and the relay to be $\eta_a$ ($\eta_b$). Denote the dark count of each detector by $p_d$.

\item
We assume the phase modulator (PM) and PBS devices at Alice and Bob are perfect.
\end{itemize}

\subsection{Photon-number channel model}
When the phases of the coherent states used by Alice and Bob are randomized, the quantum channel can be modeled as a photon-number channel model \cite{Lo:Decoy:2005}. That is, Alice and Bob randomly choose quantum channels (with a Poisson distribution) with different Fock states. Thus, the gain and QBER is composed of all the possible $i\uplus j$-channels,
\begin{equation} \label{MIFluc:Model:GainQBER}
\begin{aligned}
Q_{\mu\nu}^{w} &= \sum_{i,j} \frac{\mu^i\nu^j}{i!j!}e^{-\mu-\nu}Y_{ij}^{w}, \\
E_{\mu\nu}^{w}Q_{\mu\nu}^{w} &= \sum_{i,j} \frac{\mu^i\nu^j}{i!j!}e^{-\mu-\nu}e_{ij}^{w}Y_{ij}^{w}, \\
\end{aligned}
\end{equation}
where $w=x,z$.
%the expected photon number of the source Alice (Bob) use is $\mu$ ($\nu$); $Y_{ij}$ is the probability to get a successful partial BSM given that Alice and Bob use the $i\uplus j$-channel and $e_{ij}$ is the corresponding error rate.

In the security proof, we assume that Eve has a full control of $Y_{ij}^{w}$ and $e_{ij}^{w}$ ranging from 0 to 1.
The purpose of using decoy states is to estimate $Y_{ij}^{w}$ and $e_{ij}^{w}$, with a particular interest in $Y_{11}^{w}$ and $e_{11}^{w}$ as only the $1\uplus 1$ channel contributes to the secret key bits.
The gain and QBER, $Q_{\mu\nu}^{w}$ and $E_{\mu\nu}^{w}$, on the other hand, are observables for Alice and Bob and are used for the above estimation.

\subsection{Asymptotic case}
In this section, we present the expected values for the parameters of interest if an infinitely long key is used. These analytical results can be obtained if we assume that the system is operating under normal conditions. We emphasize that the results of this simulation model can only be used for simulation purposes, but not for the security proof. For the post-measurement processing of a real QKD experiment, the key rate and the actual key are derived from the measurement outcomes, which also include possible Eve's intervention.

Here, we directly take the results from the Appendixes of Ref.~\cite{MXF:MIQKD:2012}. The observables we need to use for the simulation are the following gains and QBERs:
\begin{equation} \label{MIFluc:Model:GainQBERsim}
\begin{aligned}
Q_{\mu\nu}^{x} &= 2y^2[1+2y^2-4yI_0(x)+I_0(2x)], \\
E_{\mu\nu}^{x}Q_{\mu\nu}^{x} &= e_0Q_{\mu\nu}^{x}-2(e_0-e_d)y^2[I_0(2x)-1], \\
\end{aligned}
\end{equation}
and
\begin{eqnarray}
\label{Path:Decoy:OriGain}
&Q_{\mu\nu}^{z} = Q_C + Q_E, & \nonumber \\
&E_{\mu\nu}^{z} Q_{\mu\nu}^{z} = e_d Q_C + (1-e_d) Q_E,&
\end{eqnarray}
where
\begin{eqnarray}
&Q_C = 2(1-p_d)^2e^{-{\mu'}/{2}} \left[1-(1-p_d)e^{-\eta_a\mu/2}\right]\left[1-(1-p_d)e^{-\eta_b\nu/2}\right],& \nonumber \\
&Q_E = 2 p_d (1-p_d)^2 e^{-{\mu'}/{2}} [I_0(2x)-(1-p_d) e^{-{\mu'}/{2}}].&
\end{eqnarray}
In the above equations, $I_0(x)$ is the modified Bessel function of the first kind, $e_d$ represents the misalignment-error probability, $e_0=1/2$, and
\begin{equation} \label{Path:Model:CoherentNotations}
\begin{aligned}
x &= \sqrt{\eta_a\mu\eta_b\nu}/2, \\
y &= (1-p_d)e^{-\mu'/4}, \\
\mu' &= \eta_a\mu+\eta_b\nu. \\
\end{aligned}
\end{equation}

We also need the gain of single-photon states, $Q_{11}^{w}$, $w=x,z$, given by
\begin{equation} \label{MIQKD:Model:Q11}
\begin{aligned}
Q_{11}^{w} = \mu\nu e^{-\mu-\nu}Y_{11}^{w}. \\
\end{aligned}
\end{equation}
Without Eve's intervention, the yield and error rate of the $1\uplus 1$ channel are given by
\begin{equation} \label{MIFluc:Model:eY11sim}
\begin{aligned}
Y_{11}^{x} = Y_{11}^{z} &=(1-p_d)^2\left[ \frac{\eta_a\eta_b}{2} +(2\eta_a+2\eta_b-3\eta_a\eta_b)p_d +4(1-\eta_a)(1-\eta_b)p_d^2 \right], \\
e_{11}^{x}Y_{11}^{x} &= e_0 Y_{11}^{x}-(e_0-e_d)(1-p_d)^2\frac{\eta_a\eta_b}{2}, \\
e_{11}^{z}Y_{11}^{z} &= e_0 Y_{11}^{z}-(e_0-e_d)(1-p_d)^2(1-2p_d)\frac{\eta_a\eta_b}{2},\\
\end{aligned}
\end{equation}
which will be used for the simulation of the asymptotic case.

\section{Post processing} \label{Sec:MIFluc:PostProcessing}

\subsection{Key rate}
The key rate is given by \cite{Lo:Decoy:2005,Lo:MIQKD:2012},
\begin{equation} \label{MIFluc:Post:KeyRate}
\begin{aligned}
R &\ge Q_{11}^{z}[1-H(e_{11}^{x})] - I_{ec}, \\
I_{ec} &= Q_{\mu\nu}^{z} f H(E_{\mu\nu}^{z}),
\end{aligned}
\end{equation}
where $I_{ec}$ is the cost of error correction, $f$ is the error correction efficiency, and $H(e)=-e\log_{2}(e)-(1-e)\log_{2}(1-e)$ is the binary Shannon entropy function. We assume that the final key is extracted from the data measured in the $z$ basis. Note that, for single-photon states, the phase error probability in the $z$ basis is the bit error probability in the $x$ basis, $e_{11}^{x}$, since single photons form a basis-independent source \cite{KoashiPreskill_03}.

\subsection{Parameter estimation} \label{Sub:MIFluc:ParaEst}
The post-measurement processing of MDI-QKD includes the two conventional stages of error correction and privacy amplification. Error correction only depends on the directly observable error rate, $E_{\mu\nu}^{z}$. Thus, the term $I_{ec}$, in the key rate formula of Eq.~\eqref{MIFluc:Post:KeyRate}, is fixed. For privacy amplification, one needs to estimate the parameters of the $1\uplus 1$-channel, $Q_{11}^{z}$ and $e_{11}^{x}$, with decoy states. Thus, the key point of the parameter estimation in this stage is to estimate the privacy amplification term, i.e., the first term on the right-hand side of Eq.~\eqref{MIFluc:Post:KeyRate}.

Assume that Alice uses $m_a$ phase-randomized coherent states with intensities $\mu_0, \mu_1, \dots, \mu_{m_a-1}$, representing one signal and $m_a-1$ decoy states, and Bob uses $m_b$ intensities $\nu_0, \nu_1, \dots, \nu_{m_b-1}$. Our objective is to solve the following \cite{MXFPhD}:
\begin{equation} \label{MIFluc:DecoyImp:minProblem}
\begin{aligned}
\min_{Y_{ij}^{w},e_{ij}^{w}} Y_{11}^{z}[1-H(e_{11}^{x})], \\
\end{aligned}
\end{equation}
subject to
\begin{equation} \label{MIFluc:DecoyImp:minConstraints}
\begin{aligned}
Q_{\mu_k\nu_l}^{w} &= \sum_{i,j} \frac{\mu_k^i\nu_l^j}{i!j!}e^{-\mu_k-\nu_l}Y_{ij}^{w}, \\
E_{\mu_k\nu_l}^{w}Q_{\mu_k\nu_l}^{w} &= \sum_{i,j} \frac{\mu_k^i\nu_l^j}{i!j!}e^{-\mu_k-\nu_l}e_{ij}^{w}Y_{ij}^{w}, \\
\end{aligned}
\end{equation}
for $k=0,1,\dots,m_a-1$, $l=0,1,\dots,m_b-1$ and $w=x,z$. The number of linear constraints in $Y_{ij}^{w}$ and $e_{ij}^{w}Y_{ij}^{w}$ is $4m_am_b$.

\begin{figure}[hbt]
\centering
\includegraphics[width = 8cm]{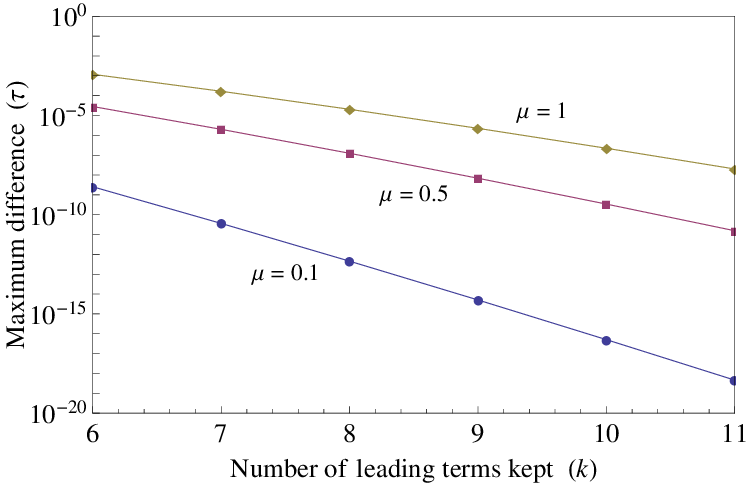}
\caption{(Color Online) Upper bounds on the dropped terms in Eq.~\eqref{MIFluc:DecoyImp:minConstraints}.  It can be seen that the effect of dropping higher order terms is negligible when the number of leading terms kept, $k$, is not too small.}
\label{Fig:dropped-terms}
\end{figure}

In order to find the minimum in Eq.~\eqref{MIFluc:DecoyImp:minProblem}, we lower bound $Y_{11}^{z}$ and upper bound  $e_{11}^{x}$ separately~\footnote{To upper bound $e_{11}^{x}$, we divide the upper bound of $e_{11}^{x}Y_{11}^{x}$ with the lower bound of $Y_{11}^{x}$.}. Both these problems can be solved using linear programming, and that will provide us with a lower bound on the optimal value that one can find by directly solving the nonlinear minimization problem in Eq.~\eqref
{MIFluc:DecoyImp:minProblem}.
Note that, even in our simplified approach, one must deal with an infinite number of unknowns in $Y_{ij}^{w}$ and $e_{ij}^{w}$, $i,j = 0,1,2, \dots$. In our numerical analysis, we take an additional simplifying step and drop terms of higher orders in Eq.~\eqref{MIFluc:DecoyImp:minConstraints}. Because of the Poisson-distributed
coefficients of $Y_{ij}^{w}$ and $e_{ij}^{w}$, in Eq.~\eqref{MIFluc:DecoyImp:minConstraints}, these terms decrease exponentially by increasing $i$ and $j$.
%This optimization problem can be solved efficiently.
From our numerical simulations, we find that the effect of terms with $i,j \ge 7$ on the parameter estimation is negligible.
To further verify this analytically, note that
the sum of the dropped terms of $i,j \ge k$ in Eq.~\eqref{MIFluc:DecoyImp:minConstraints} is upper bounded by $\tau(\mu,k):=1-(\sum_{i=0}^{k-1} \frac{\mu^i}{i!}e^{-\mu})^2$ when considering  $Y_{ij}^{w}=1$ and assuming that $\mu=\mu_k=\mu_l$.
Fig.~\ref{Fig:dropped-terms} shows $\tau(\mu,k)$ for three nominal values of $\mu$ and $k=6,\ldots,11$. It turns out that the neglected terms have insignificant impact on the values of $Y_{11}^{w}$ and $e_{11}^{w}Y_{11}^{w}$ that we obtain in our simulations in Sec.~\ref{Sec:MIfluc:Sim}; see Tables~\ref{Tab:MIFluc:eY11estVW} and \ref{Tab:MIFluc:eY11est4s}.
%As we show later in Sec.~\ref{Sec:MIfluc:Sim}, these numbers $\tau(\mu,k)$ are insignificant compared to $Y_{11}^{w}$ and $e_{11}^{w}Y_{11}^{w}$ obtained in the simulations (see Tables~\ref{Tab:MIFluc:eY11estVW} and \ref{Tab:MIFluc:eY11est4s}).
%Table~\ref{Tab:dropped-terms} shows that this bound is small and negligible for typical $\mu$ and $k=6,10$.

%\begin{table}[hbt]
%\centering
%\begin{tabular}{ccccc}
%\hline
%$\mu$ & $\tau(\mu,6)$ & $\tau(\mu,10)$ \\

%$\mu$ & $\sum_{i=6}^\infty \frac{\mu^i}{i!}e^{-\mu}$ & $\sum_{i=10}^\infty \frac{\mu^i}{i!}e^{-\mu}$ \\
%\hline
%$0.1$ & $3.6 \times 10^{-11}$ & $4.6 \times 10^{-19}$\\

%$0.1$ & $1.3 \times 10^{-9}$ & $2.5 \times 10^{-17}$\\
%\hline
%$0.5$ & $2.0 \times 10^{-6}$ & $1.5 \times 10^{-11}$\\

%$0.5$ & $1.4 \times 10^{-5}$ & $1.7 \times 10^{-10}$\\
%\hline
%$1$ & $1.6 \times 10^{-4}$ & $2.0 \times 10^{-8}$\\

%$1$ & $5.9 \times 10^{-4}$ & $1.1 \times 10^{-7}$\\
%\hline
%\end{tabular}
%\caption{Upper bounds of the dropped terms in Eq.~\eqref{MIFluc:DecoyImp:minConstraints}.  It can be seen that the effect of dropping higher order terms is negligible.} \label{Tab:dropped-terms}
%\end{table}

We follow the statistical fluctuation analysis proposed in Ref.~\cite{MXF:Practical:2005}. Then, the equalities in Eq.~\eqref{MIFluc:DecoyImp:minConstraints} becomes inequalities,
\begin{equation} \label{MIFluc:DecoyImp:ConstraintsIneq}
\begin{aligned}
\hat{Q}_{\mu_k\nu_l}^{w}(1-\beta_q) &\le \sum_{i,j} \frac{\mu_k^i\nu_l^j}{i!j!}e^{-\mu_k-\nu_l}Y_{ij}^{w} \le \hat{Q}_{\mu_k\nu_l}^{w}(1+\beta_q) \\
\hat{E}_{\mu_k\nu_l}^{w}\hat{Q}_{\mu_k\nu_l}^{w}(1-\beta_{eq}) &\le \sum_{i,j} \frac{\mu_k^i\nu_l^j}{i!j!}e^{-\mu_k-\nu_l}e_{ij}^{w}Y_{ij}^{w} \le \hat{E}_{\mu_k\nu_l}^{w}\hat{Q}_{\mu_k\nu_l}^{w}(1+\beta_{eq}), \\
\end{aligned}
\end{equation}
where if the left hand side of the inequality is negative, we replace it with 0. The variables, $\hat{Q}_{\mu_k\nu_l}^{w}$ and $\hat{E}_{\mu_k\nu_l}^{w}$ are measurement outcomes. That is, they are rates instead of probabilities. The fluctuation ratio $\beta_q$ and $\beta_{eq}$ can be evaluated by
\begin{equation}
%\label{MIFluc:DecoyImp:ConstraintsIneq}
\begin{aligned}
\beta_q &= \frac{n_{\alpha}}{\sqrt{N_{\mu_k\nu_l}^{w}\hat{Q}_{\mu_k\nu_l}^{w}}}, \\
\beta_{eq} &= \frac{n_{\alpha}}{\sqrt{N_{\mu_k\nu_l}^{w}\hat{E}_{\mu_k\nu_l}^{w}\hat{Q}_{\mu_k\nu_l}^{w}}}, \\
\end{aligned}
\end{equation}
where $N_{\mu_k\nu_l}^{w}$ is the number of pulses, in the $w$ basis, sent out by Alice and Bob when they use intensities $\mu_k$ and $\nu_l$, respectively; $n_{\alpha}$ is the number of standard deviations one chooses for statistical fluctuation analysis. In other words, $N_{\mu_k\nu_l}^{w}\hat{Q}_{\mu_k\nu_l}^{w}$ is the number of successful partial BSMs when Alice and Bob use intensities $\mu_k$ and $\nu_l$, respectively, and $N_{\mu_k\nu_l}^{w}\hat{E}_{\mu_k\nu_l}^{w}\hat{Q}_{\mu_k\nu_l}^{w}$ is the corresponding error count. If we follow the Gaussian assumption made in \cite{MXF:Practical:2005}, the number of standard deviations, $n_\alpha$, will be directly related to the failure probability of this security analysis. For example, when $n_\alpha=5$, as used later, it will introduce a failure probability of $5.73\times10^{-7}$.

\section{Simulation} \label{Sec:MIfluc:Sim}
For simplicity, we assume that Alice and Bob send the same number of pulses for all $\mu\uplus\nu$ channels, denoted by $N_{data}$. In the following simulations, the parameters of the experimental setup are listed in Table~\ref{Tab:MIFluc:SetupPara}.
\begin{table}[hbt]
\centering
\begin{tabular}{ccccccc}
\hline
$e_d$ & $p_d$ & $f$ & $N_{data}$ \\
\hline
1.5\% & $3\times10^{-6}$ & 1.16 & $2\times10^{10}$ \\
\hline
\end{tabular}
\caption{List of experimental parameters used in numerical results: $e_d$ is the misalignment probability; $p_d$ is the background count rate per detector; $f$ is the error correction inefficiency; $N_{data}$ is the number of pulses sent by Alice and Bob for each pair of intensities.} \label{Tab:MIFluc:SetupPara}
\end{table}

In the simulation, we absorb the detection loss into channel losses. Note that with the current development in high-speed QKD systems \cite{Yuan:Selfdif:2007,Sasaki:TokyoQKD:2011,Wang:260kmQKD:2012}, $N_{data}=2\times10^{10}$ pulses can be transmitted in seconds. We assume that Alice and Bob pick $n_{\alpha}$ standard deviations for the statistical fluctuation analysis, which is determined by the allowable failure probability for the system.

\subsection{Vacuum+weak-decoy-state protocol} \label{Sub:MIFluc:VW}
We consider that Alice and Bob run the vacuum+weak-decoy-state protocol \cite{MXF:Practical:2005} and they choose the same intensities for the coherent states. Let us assume a typical set of intensities: $\{0,0.1,0.5\}$. Note that we assume $N_{data}=2\times10^{10}$ for each $\mu\uplus\nu$ channel. Thus, the total number of pulses sent by Alice and Bob is $18\times10^{10}$.

For each of the nine $\mu\uplus\nu$ channels, Alice and Bob can obtain a set of linear inequalities, in the form of Eq.~\eqref{MIFluc:DecoyImp:ConstraintsIneq}, for gains and QBERs. As noted before, we neglect terms with $i,j \ge 7$, and find the lower bound on $Y_{11}^{z}$ and the upper bound on $e_{11}^{x}$ using linear programming.
%It is not hard to see
%We remark
%that the lower bound of $Y_{11}$ and upper bound of $e_{11}Y_{11}$ can be efficiently solved by linear programming.

In order to obtain a sense of the magnitude of the parameter values, we calculate the gains and QBERs, at $\eta_a=\eta_b=0.1$, using Eqs.~\eqref{MIFluc:Model:GainQBERsim} and \eqref{Path:Decoy:OriGain} for the $x$ and $z$ basis. The gain values are listed in Table~\ref{Tab:MIFluc:9Gains}.
\begin{table}[hbt]
\centering
\begin{tabular}{c|ccc|ccc}
\hline
& \multicolumn{3}{|c|}{z-basis} & \multicolumn{3}{|c}{x-basis} \\
\hline
Bob/Alice & $\mu=0$ & $\mu=0.1$ & $\mu=0.5$ & $\mu=0$ & $\mu=0.1$ & $\mu=0.5$ \\
\hline
$\nu=0$ & $3.6000\times10^{-11}$ & $5.9587\times10^{-8}$ & $2.8900\times10^{-7}$ & $3.5999\times10^{-11}$ & $2.4873\times10^{-5}$ & $6.0229\times10^{-4}$ \\
$\nu=0.1$ & $5.9587\times10^{-8}$ & $4.9374\times10^{-5}$ & $2.3935\times10^{-4}$ & $2.4873\times10^{-5}$ & $9.8876\times10^{-5}$ & $8.6437\times10^{-4}$ \\
$\nu=0.5$ & $2.8900\times10^{-7}$ & $2.3935\times10^{-4}$ & $1.1603\times10^{-3}$ & $6.0229\times10^{-4}$ & $8.6437\times10^{-4}$ & $2.3495\times10^{-3}$ \\
\hline
\end{tabular}
\caption{Simulation values of the gain, $Q_{\mu \nu}^{w}$, $w=x,z$, for the vacuum+weak-decoy-state protocol, evaluated using Eqs.~\eqref{MIFluc:Model:GainQBERsim} and \eqref{Path:Decoy:OriGain}. It is assumed that $\eta_a=\eta_b=0.1$.} \label{Tab:MIFluc:9Gains}
\end{table}
The QBER of the case when either party chooses the vacuum decoy state is 1/2 and that of the remaining four nontrivial cases is shown in Table~\ref{Tab:MIFluc:4QBERs}, for the $x$ and $z$ basis. Note that the QBER in the $z$ basis is reasonably close to $e_d$ as expected from Eqs.~\eqref{Path:Decoy:OriGain} and \eqref{MIFluc:Model:eY11sim}. The QBER in the $x$ basis, on the other hand, is larger than 25\%, which is mainly caused by false triggering of multiphoton states \cite{Lo:MIQKD:2012,MXF:MIQKD:2012}. This is the key reason why the final key in Eq.~\eqref{MIFluc:Post:KeyRate} should be only extracted from the $z$ basis.

\begin{table}[hbt]
\centering
\begin{tabular}{c|cc|cc}
\hline
& \multicolumn{2}{|c|}{z-basis} & \multicolumn{2}{|c}{x-basis} \\
\hline
Bob/Alice & $\mu=0.1$ & $\mu=0.5$ & $\mu=0.1$ & $\mu=0.5$ \\
\hline
$\nu=0.1$ & $1.6164\%$ & $1.5700\%$ & $25.7184\%$ & $36.3867\%$ \\
$\nu=0.5$ & $1.5700\%$ & $1.5236\%$ & $36.3867\%$ & $25.4516\%$ \\
\hline
\end{tabular}
\caption{Simulation values of QBER, $E_{\mu \nu}^{w}$, $w=x,z$, for different intensity values evaluated from Eqs.~\eqref{MIFluc:Model:GainQBERsim} and \eqref{Path:Decoy:OriGain}.} \label{Tab:MIFluc:4QBERs}
\end{table}

%In practice, Table \ref{Tab:MIFluc:9Gains} and \ref{Tab:MIFluc:4QBERs} are given by the raw data from an experiment. The task of the security analysis, the main focus of this work, is to determine the final secure key rate for this set of data.
%%Following the analysis given in Section \ref{Sub:MIFluc:ParaEst}, we can minimize $Y_{11}^{z}$ and maximize $e_{11}^{x}$ with constraints of Eq.~\eqref{MIFluc:DecoyImp:ConstraintsIneq}, by substituting the simulation values listed in Table \ref{Tab:MIFluc:9Gains} and \ref{Tab:MIFluc:4QBERs}.
%Following the analysis given in Section \ref{Sub:MIFluc:ParaEst}, we minimize $Y_{11}^{z}$ and maximize $e_{11}^{x}$ subject to constraints of Eq.~\eqref{MIFluc:DecoyImp:ConstraintsIneq}, by assuming that the values listed in Table \ref{Tab:MIFluc:9Gains} and \ref{Tab:MIFluc:4QBERs} are estimated values for, respectively, gain and QBER, in a certain experiment.

In practice, $\hat Q_{\mu \nu}^{w}$ and $\hat E_{\mu \nu}^{w}$, similar to those in Tables \ref{Tab:MIFluc:9Gains} and \ref{Tab:MIFluc:4QBERs}, are derived from the raw data obtained in the experiment. The task of the security analysis, which is the main focus of this work, is to determine the final secure key rate for such sets of data.  Following the analysis given in Sec.~\ref{Sub:MIFluc:ParaEst}, here we minimize $Y_{11}^{z}$ and maximize $e_{11}^{x}$ subject to constraints of Eq.~\eqref{MIFluc:DecoyImp:ConstraintsIneq}, by assuming that the values listed in Tables \ref{Tab:MIFluc:9Gains} and \ref{Tab:MIFluc:4QBERs} are, respectively, the measured gain and QBER in a certain experiment.

Table \ref{Tab:MIFluc:eY11estVW} provides lower and upper bounds on these parameters, obtained by solving the corresponding linear-programming problems, and compared them with the expected values from the simulation results of Eq.~\eqref{MIFluc:Model:eY11sim}. From Table~\ref{Tab:MIFluc:eY11estVW}, one can see that the parameter estimations in the $x$ basis is worse than those in the $z$ basis. This is because multiphoton terms contribute more to the gains and QBERs in the $x$ basis than in the $z$ basis.

\begin{table}[hbt]
\centering
\begin{tabular}{c|c|cc|cc}
\hline
& & \multicolumn{2}{|c|}{$w=z$} & \multicolumn{2}{|c}{$w=x$} \\
\hline
Parameters & Asymptotic value & Lower bound & Upper bound & Lower bound & Upper bound \\
\hline
$Y_{11}^{w}$ & $5.0011\times10^{-3}$ & $4.6043\times10^{-3}$ & $6.0286\times10^{-3}$ & $4.1343\times10^{-3}$ & $6.6334\times10^{-3}$ \\
$e_{11}^{w}$ & $1.5108\%$ & $0.9556\%$ & $2.1341\%$ & $0$ & $10.2126\%$ \\
\hline
\end{tabular}
\caption{Lower and upper bounds on $Y_{11}^{w}$ and $e_{11}^{w}$ in both bases, compared to asymptotic values. In our statistical fluctuation analysis, five standard deviations are taken into consideration.} \label{Tab:MIFluc:eY11estVW}
\end{table}

\begin{figure}[hbt]
\centering
\resizebox{12cm}{!}{\includegraphics{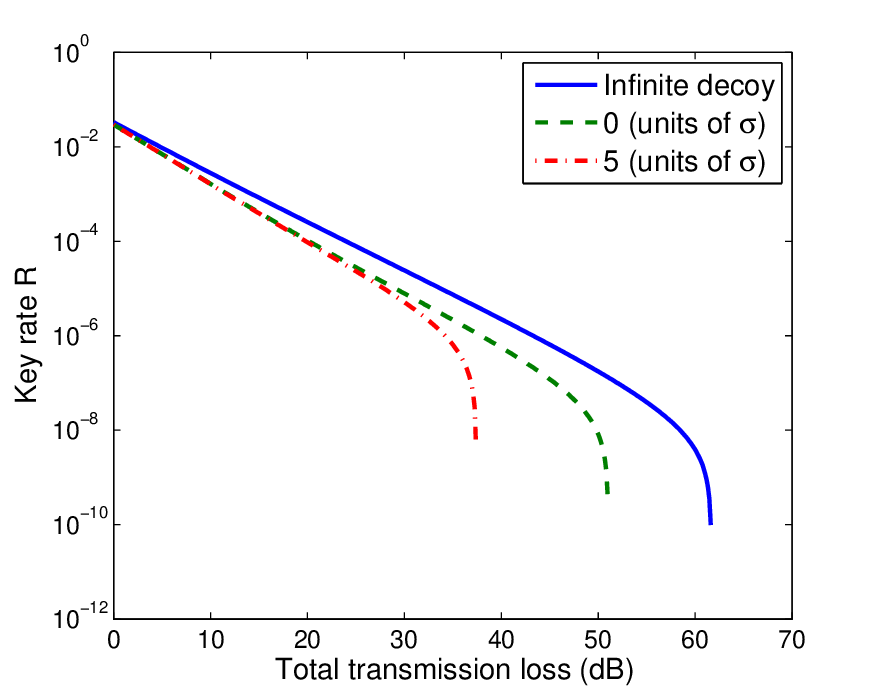}}
\caption{Key rate versus channel transmittance using vacuum+weak decoy-state method for MDI-QKD.}
\label{Fig:MIFluc:RdBVW}
\end{figure}

Substituting the parameter estimations from Table \ref{Tab:MIFluc:eY11estVW}, the lower bound of $Y_{11}^z$ and the upper bound of $e_{11}^x$, into Eq.~\eqref{MIFluc:Post:KeyRate}, one can calculate the key rate to be $6.89\times10^{-5}$ bits/pulse. Similarly, one can evaluate the dependence of the key rate on channel transmittance, as shown in Fig.~\ref{Fig:MIFluc:RdBVW}. One can see that even by including statistical fluctuations the key rate decreases linearly with channel loss before the cut-off regime. In the low-loss regime, the vacuum+weak-decoy-state protocol performs almost as well as the asymptotic case.

As shown in Fig.~\ref{Fig:MIFluc:RdBVW}, with $n_{\alpha}=5$ standard deviations, the maximum tolerable transmission loss is almost 30~dB less than that of the asymptotic case. Even if we do not take the statistical fluctuations ($n_{\alpha}=0$) into account, there is still a gap between the two cases.
%This gap is expected as one can see from Table~\ref{Tab:MIFluc:eY11estVW} that the estimation of $Y_{11}$ and $e_{11}$ does not perform well when $\eta_a=\eta_b=0.1$, corresponding to 20 dB total loss \textcolor{red}{(I'm not convinced with this explanation. The table is for 5 standard deviations, and we cannot extend its implications to the case of $n_{\alpha}=0$. What do lower and upper bounds for $Y_{11}$ mean when in effect we have an equality, rather than two inequalities, in our linear programming?)}.
Thus, there is big room for further improvement. In the next simulation, we will consider three decoy states and show that further improvements can be made when more decoy states are applied.

\subsection{Vacuum+two-weak-decoy-state protocol} \label{Sub:MIFluc:4decoy}
In order to give a better estimation of $Y_{11}^{z}$ and $e_{11}^{x}$, one can use more than two decoy states. Let us assume that Alice and Bob use four coherent states $\{0,0.1,0.2,0.5\}$ and that we use $N_{data}=2\times10^{10}$ for each $\mu\uplus\nu$ channel. Thus, the total number of pulses sent by Alice and Bob is $32\times10^{10}$, corresponding to 16 channels. Given that by adding an extra decoy state on each side we can better estimate channel parameters, the key rate is expected to be no less than the one in Sec.~\ref{Sub:MIFluc:VW}.

Similar to the previous section, we take $\eta_a=\eta_b=0.1$ as an example to see how accurate the parameter estimation is. The bounds of $Y_{11}^{z}$ and $e_{11}^{x}$ in both bases, compared to the asymptotic case, are listed in Table~\ref{Tab:MIFluc:eY11est4s}. Again, the parameter estimations in the $x$ basis is worse than those in the $z$ basis, due to the multiphoton terms.

\begin{table}[hbt]
\centering
\begin{tabular}{c|c|cc|cc}
\hline
& & \multicolumn{2}{|c|}{$w=z$} & \multicolumn{2}{|c}{$w=x$} \\
\hline
Parameters & Asymptotic value & Lower bound & Upper bound & Lower bound & Upper bound \\
\hline
$Y_{11}^{w}$ & $5.0011\times10^{-3}$ & $4.7058\times10^{-3}$ & $5.2377\times10^{-3}$ & $4.3734\times10^{-3}$ & $5.5640\times10^{-3}$ \\
$e_{11}^{w}$ & $1.5108\%$ & $1.1103\%$ & $2.0409\%$ & $0$ & $7.7954\%$ \\
\hline
\end{tabular}
\caption{Lower and upper bounds on $Y_{11}^{w}$ and $e_{11}^{w}$ in both bases, compared to asymptotic values. In our statistical fluctuation analysis, five standard deviations are taken into consideration.} \label{Tab:MIFluc:eY11est4s}
\end{table}

%Similar to the vacuum+weak-decoy-state case, one can calculate the key rate by substituting the parameter estimations from Table \ref{Tab:MIFluc:eY11est4s}, the lower bound of $Y_{11}^z$ and the upper bound of $e_{11}^x$, into Eq.~\eqref{MIFluc:Post:KeyRate}, one can calculate the key rate to be $1.09\times10^{-4}$ bits/pulse.

Similar to the vacuum+weak-decoy-state case, one can calculate the key rate to be $1.09 \times 10^{-4}$ bits/pulse by substituting the parameter estimations from Table \ref{Tab:MIFluc:eY11est4s}, the lower bound of $Y^z_{11}$ and the upper bound of $e^x_{11}$, into Eq.~\eqref{MIFluc:Post:KeyRate}. According to Table \ref{Tab:MIFluc:eY11est4s}, our parameter estimation has improved when more decoy states (in extra pulses) are applied, as compared to the previous case in Table \ref{Tab:MIFluc:eY11estVW}.

\begin{figure}[hbt]
\centering
\resizebox{12cm}{!}{\includegraphics{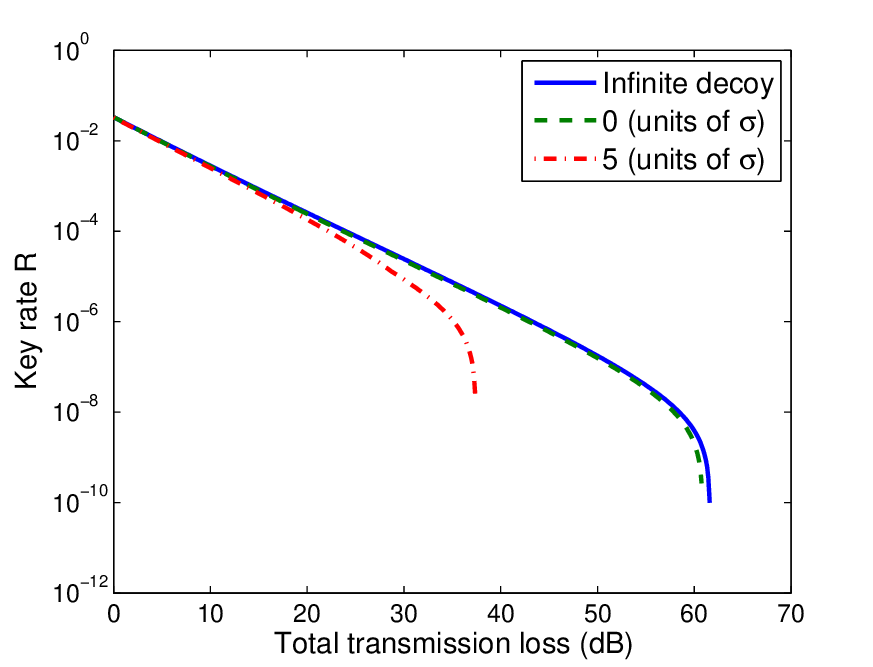}}
\caption{Key rate versus channel transmittance using vacuum+two-weak-decoy-state method for MDI-QKD.}
\label{Fig:MIFluc:RdB4s}
\end{figure}

The dependence of the key rate on the channel transmittance is shown in Fig.~\ref{Fig:MIFluc:RdB4s}. One can see that the gap between the finite-size case and the asymptotic case is smaller than the one shown in Fig.~\ref{Fig:MIFluc:RdBVW}. In the case of $n_{\alpha}=0$, the vacuum+two-weak-decoy-state protocol is very close to the asymptotic case. This is different from regular decoy-state protocol, where two decoy states are proven to be sufficient for practical usage \cite{MXF:Practical:2005}.

\section{Conclusions} \label{Sec:MIFluc:Conclusion}
We showed that MDI-QKD is a highly practical scheme even when the statistical fluctuations are accounted for. In the low-loss regime, with only two or three decoy states, the performance of MDI-QKD with statistical fluctuations is close to that of the asymptotic case. At higher values of loss, using three decoy states would be recommended.
We remark that our analysis is quite general and is applicable to different MDI-QKD implementations such as those based on phase encoding and/or polarization encoding as well as those those based on the BB84 protocol or the six-state protocol.

%%%%%%%%%%%%%%%%%%%%%%%%%%%%%%%%%%%%%%%%%%%%%%%%%%%%%%%%%%%%%%%%%%%
% Acknowledgments
\section*{Acknowledgments}
The authors would like to thank H.~-K.~Lo, T.~F.~da Silva and H.~-L.~Yin for enlightening discussions and for his help in the preparation of Figure \ref{Fig:Path:MIdiag}. The authors gratefully acknowledge the financial support from National Basic Research Program of China Grants No.~2011CBA00300 and No.~2011CBA00301, National Natural Science Foundation of China Grants No.~61073174, No.~61033001, and No.~61061130540, the 1000 Youth Fellowship program in China, the European Community's Seventh Framework Programme under Grant Agreement 277110, the UK Engineering and Physical Science Research Council (Grant No.~EP/J005762/1), and Hong Kong RGC Grant No.~700709P.

%\begin{appendix}
%\section{Problem with previous analysis} \label{App:MIQKD:6state}
%
%the previous analysis given by practical decoy \cite{MXF:Practical:2005} is not working here, as used by Wang.
%
%$e_{11}$ will be larger than 50\%. First Emu is around 20\% and the ratio of $Q_{11}$ over Qmu is less than around 10\%. then $e11>50\%$.
%
%\end{appendix}

%%%%%%%%%%%%%%%%%%%%%%%%%%%%%%%%%%%%%%%%
% choose a style
%\bibliographystyle{ieeetr}
%\bibliographystyle{unsrt}
\bibliographystyle{apsrev4-1}
%%%%%%%%%%%%%%%%%%%%%%%%%%%%%%%%%%%%%%%%

%%%%%%%%%%%%%%%%%%%%%%%%%%%%%%%%%%%%%%%%
% choose a .bib file
\bibliography{Bibli}
%%%%%%%%%%%%%%%%%%%%%%%%%%%%%%%%%%%%%%%%

%\nocite{*}

\end{document}